\documentclass[onecolumn,secnumarabic,amssymb, nobibnotes, aps, prd]{revtex4}

\usepackage{graphicx}
\usepackage{epsfig}

\def\ket#1{\mathinner{|{#1}\rangle}}

\begin{document}

\title{Deterministic generation of non-classical states of light using photon blockade}%

\author{Andrei Faraon$^a$, Arka Majumdar$^a$, Jelena Vuckovic}%
\email[Correspondence should be addressed to A.F.: ]{faraon@stanford.edu}
\affiliation{E.L.Ginzton Laboratory, Stanford University, Stanford, CA, 94305\\
$^a$ These authors contributed equally to this work}

\begin{abstract}

The generation of non-classical states of light via photon blockade with time-modulated input is analyzed. We show that improved single photon statistics can be obtained by adequately choosing the parameters of the driving laser pulses. An alternative method, where the system is driven via a continuous wave laser and the frequency of the dipole is controlled (e.g. electrically) at very fast timescales is presented.

\end{abstract}

\maketitle

\section{ {Introduction}}

Photons at optical frequencies are commonly used in quantum key distribution systems\cite{BB84}, and play a fundamental role in proposed devices for quantum information processing\cite{DuanKimble04,KLM01}. While the majority of quantum cryptography systems currently in use are based on coherent light sources, most of the proposals for more advanced quantum information devices are based on non-classical states of light, mainly single photon states. These type of devices include quantum repeaters \cite{ChildressLukin,2006.Ladd.HybridQuantumRepeater}, linear optics quantum computing devices\cite{KLM01}, and quantum networks based on cavity quantum electrodynamics\cite{CZKM1997PRL}. Considerable research has been done to make deterministic single photon sources, which can generate indistinguishable single photons with very high efficiency. One way to generate single photons is by parametric down conversion or attenuation of coherent laser beams such that the probability of having multi-photon states is considerably diminished. However, this causes increased probability of the vacuum state, so most of the time there are no photons in the light source. To make a good single photon source one has to ensure suppression of both the multi-photon states as well as the zero-photon state. There are several proposals for making on demand single photon sources, like above-band or resonant excitation of solid state single emitters \cite{VuckovicSingPhotDem} or adiabatic state transfer in single atoms \cite{2007.Rempe.SinglePhotServ}.

In the recent years, there has been increased interest in generating non-classical states of light via strongly coupled cavity quantum electrodynamics (CQED) systems\cite{97ImamogluBlockade,2008.PRL.Koch.PhotonStatistics,2008.NatPhys.Rempe.NonlinearSpec,2009.PRL.Rempe.TwoPhoton}. Non-classical states of light that closely resemble the single photon states can be generated in this kind of systems by operating in the photon blockade regime\cite{97ImamogluBlockade} with a single emitter strongly coupled to an optical resonator. The photon blockade was first observed in atomic physics \cite{KimbleBlockade} and recently has been demonstrated in solid state CQED using quantum dots in photonic crystal cavities\cite{BlockadeTunnelingPaper}. These experiments demonstrated photon anti-bunching caused by photon blockade, a signature of enhanced probability of the single photon states at the output. There are a few advantages for using photon blockade for single photon generation. First, the produced single photons are free of jitter. Second, the collection efficiency of the photons is very high due to the presence of a cavity. However, for on demand single photon generation, the source must be operated in pulsed regime and one photon should be generated for each pulse. This requirement was not satisfied in any experiments performed so far on photon blockade. 

In this paper, we analyze the optimal conditions that must be satisfied by the CQED system for on demand generation of single photon states via photon blockade.  The system considered here is a quantum dot (QD) strongly coupled to a photonic crystal cavity, although our conclusion is valid for any cavity QED system. We analyze two different driving configurations that lead to on demand single photon generation. In section \ref{sec:pulse_shaping} we consider the system coherently driven by laser pulses. In section \ref{sec:stark_shift_operation} we consider a system driven by a continuous wave laser where the deterministic single photon generation is done by electrical control of the QD resonance frequency.

\section{Photon blockade in pulsed operation of the driving field} 
\label{sec:pulse_shaping}

\begin{figure}[htp]
\includegraphics[width=3.5in]{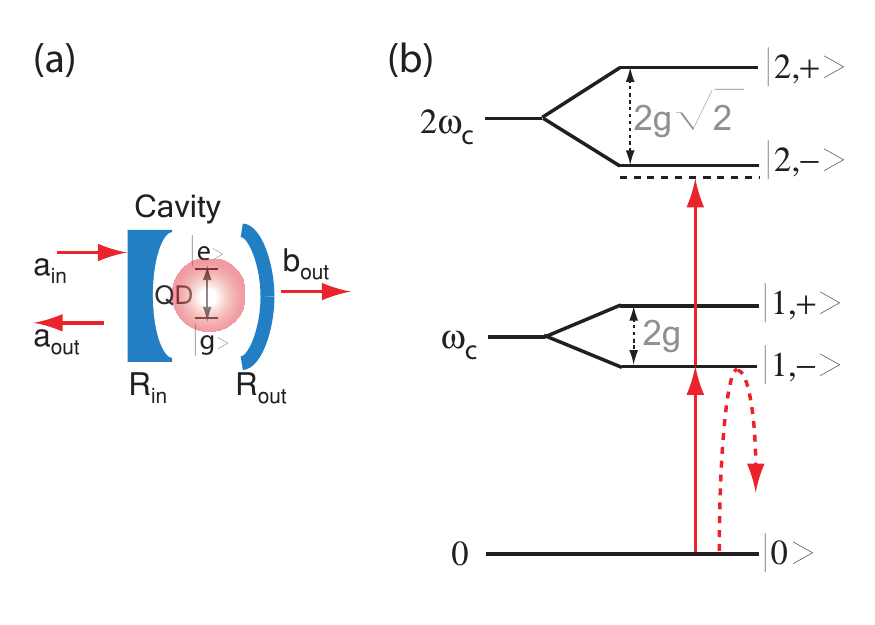}
\caption{(a) Schematic representation of a single emitter coupled to an optical resonator. The optical resonator has mirrors with different reflection coefficients ($R_{in}>>R_{out}$) such that most of the field coupled into the cavity is emitted at the output port. (b) Schematic representation of the energy eigenstates for a strongly coupled cavity-emitter system.} 
\label{Fig_Photon_Blockade}
\end{figure}

The studied optical system consists of a single emitter (QD) strongly coupled to a cavity, as shown in Fig.\ref{Fig_Photon_Blockade}. When coherently driven by a laser field, the strongly coupled system is well described by Jaynes-Cummings Hamiltonian:

\begin{equation}
\label{eqn:H}
H=\hbar\left[\omega_a \sigma_+ \sigma_-+\omega_ca^\dag
a+ig(a^\dag\sigma_--a\sigma_+)+\Omega(t)(ae^{i\omega_lt}+a^\dag e^{-i\omega_lt})\right]
\end{equation}

\noindent where, $\omega_a$ and $\omega_c$  are the QD and the cavity resonance frequency respectively and $\omega_l$ is the driving laser frequency, $g$ is the coupling strength of the QD to the cavity mode. $\Omega(t)$ is the Rabi frequency of the laser and is given by $\Omega(t)=\mu E(t)/\hbar$, where $\mu$ is the dipole moment of the transition being driven by the laser and $E(t)$ is the temporal electric field of the laser, inside the cavity; $a$ is the annihilation operator for the cavity mode. If the excited and ground state of the QD are denoted by $|e\rangle$ and $|g\rangle$ then $\sigma_-=|g\rangle\langle e|$, $\sigma_+=|e\rangle\langle g|$. In a frame rotating in laser frequency, under Rotating Wave Approximation (RWA) the Hamiltonian can be written as:

\begin{equation}
\label{eqn:H2} H=\hbar\left[\Delta_a \sigma_+ \sigma_-+\Delta_c
a^\dag a+ig(a^\dag\sigma_--a\sigma_+)+\Omega(t)(a+a^\dag )\right]
\end{equation}

\noindent where $\Delta_a=\omega_a-\omega_l$ and $\Delta_c=\omega_c-\omega_l$ are the detuning of the QD and the cavity with the laser. To fully describe a real system, loss must be included in the Hamiltonian. The two main loss mechanisms are the cavity field decay rate $\kappa=\omega_{c}/2Q$ ($Q$ is the quality factor of the resonator) and QD spontaneous emission rate $\gamma$. When the coupling strength $g$ is greater than the loss rates $\kappa$ and $\gamma$, the system is in the strong coupling regime \cite{AndreiTtune,NatureRef}. In this regime, energy eigenstates are grouped in two-level manifolds with eigen-energies given by $n\omega_c \pm g\sqrt{n}$ (for $\omega_a=\omega_c$), where $n$ is the number of energy quanta in the cavity-QD system. The eigenstates can be written as:

\begin{eqnarray}
|n,+\rangle=\frac{|g,n\rangle + |e,n-1\rangle}{\sqrt{2}}\\
|n,-\rangle=\frac{|g,n\rangle - |e,n-1\rangle}{\sqrt{2}}
\end{eqnarray}

The splitting between the energy eigenstates in each manifold has a non-linear dependence on $n$. This anharmonicity in the splitting of the energy eigenstates gives rise to nonlinear optics phenomena at single photon level. One of these phenomena is photon blockade, where the presence of one photon in the cavity blocks the coupling of subsequent photons. For example, the system could be driven by a coherent light source (see Fig.\ref{Fig_Photon_Blockade}) with frequency resonant with one of the polaritons (say $|1,-\rangle$). Once a photon is coupled, the system is excited into the state $|1,-\rangle$, so the coupling of another photon with energy $\omega_c - g$ would require the system to transition to energy state $2(\omega_c - g)$. However, the system does not have an eigenstate at this energy, the closest being at $2\omega_c - g\sqrt{2}$. For this reason, the probability of coupling the second photon is reduced.

A typical experimental configuration for observing photon blockade is depicted in Fig.\ref{Fig_Photon_Blockade}(a) where light can be coupled into the cavity via the input port $a_{in}$ and the output can be collected at the port $b_{out}$. The resonator is represented by two mirrors with reflection coefficients $R_{in}$ and $R_{out}$.

\begin{figure*}
\centering
\includegraphics[width=5in]{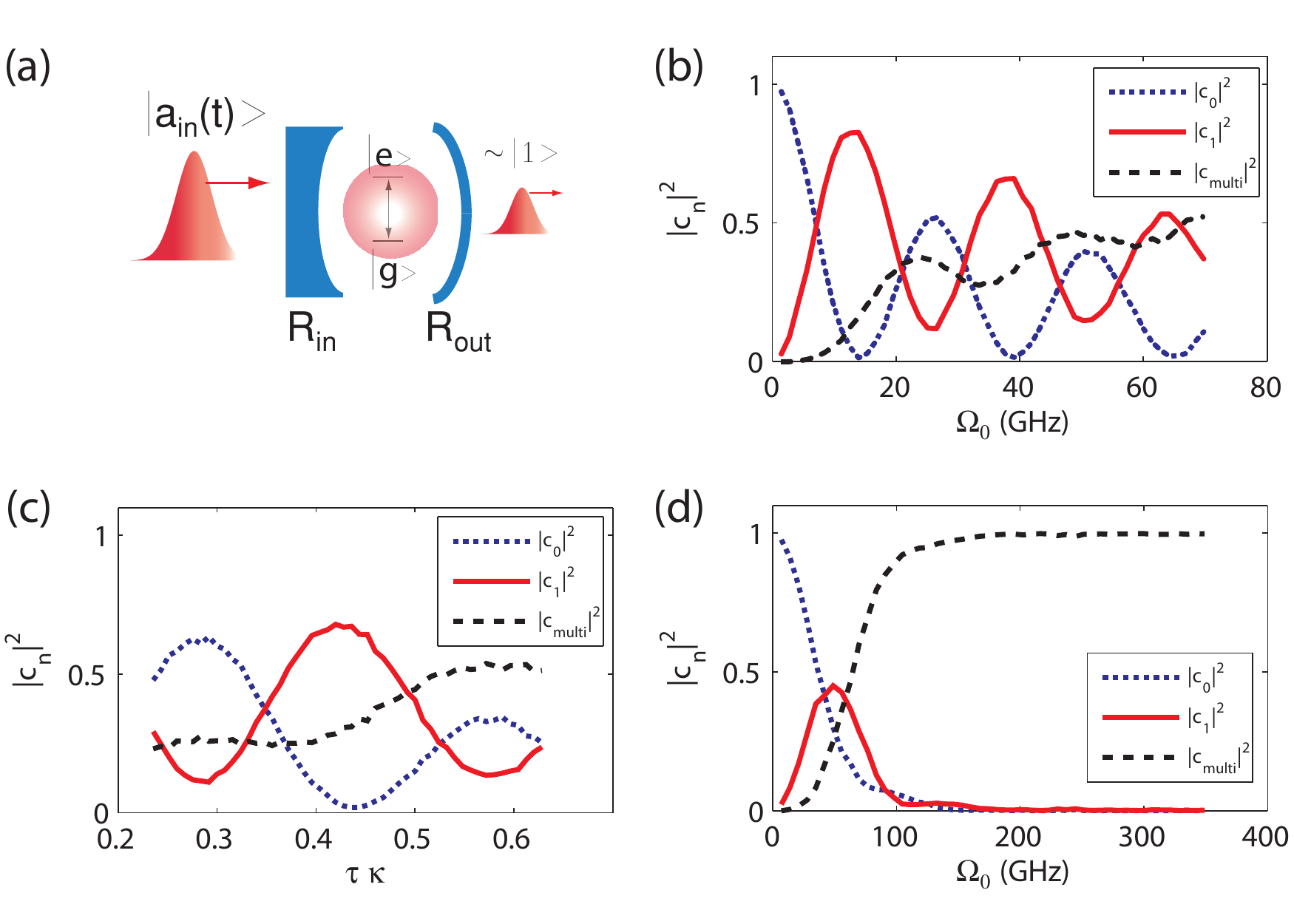}
\caption{(a) Non-classical state generation via pulsed-operation in photon blockade. The properties of the laser pulse coupled at the input port are controlled such that the output field has primarily a single-photon component. (b) Normalized Fock-state coefficients ($|c_{0}|^{2}$ for vacuum state, $|c_{1}|^{2}$ for single photon state, and $|c_{multi}|^{2}=\sum^{\infty}_{n=2}|c_{n}|^{2}$ for multi photon probability) for the output field as the intensity of the laser pulse $\Omega_{0}$ is modified and the pulsewidth is kept constant at $\tau = 0.45/\kappa$. The ground state and one of the first order eigenstates form an effective two level system so Rabi oscillations are observed in the single photon character of the output field. The system has parameters $\kappa/2\pi=1GHz$, $\gamma/2\pi=0.1GHz$, $g/2\pi=40GHz$. (c) Fock-state coefficients for the output field as the duration of the laser pulse ($\tau$) is modified while $\Omega_{0}= 10 GHz$ was kept constant. The system has parameters $\kappa/2\pi=1GHz$, $\gamma/2\pi=0.1GHz$, $g/2\pi=40GHz$. (d) The same simulation as in the panel {\bf b} for a system with $\kappa/2\pi=5GHz$, $g/2\pi=30GHz$} 
\label{Fig_shaping}
\end{figure*}

For on demand generation of non-classical states of light, it is desirable that the state is delivered at the output port at times that can be chosen deterministically. This means that the interaction between the input beam and the cavity/QD system is controlled in time domain. One way to achieve this is by driving the system with light pulses with controlled shape, as illustrated schematically in Fig.\ref{Fig_shaping}(a). In this section, we investigate how the non-classical state of light observed at the output depends on the properties of the laser pulse at the input.

In principle, the input laser pulse can have any specific waveform. To limit the number of free parameters, we analyze the configuration where the system is driven with Gaussian pulses. In this case, the free parameters are the length, intensity and the center frequency of the pulse.

To simulate the system we use Quantum Trajectory method \cite{CarmichaelOpenSystems}. The evolution of the density matrix $\rho$ of the system, including all the losses is given by (in the normalized form with $\hbar =1 $):
\begin{equation}
\label{Maseq} \frac{d\rho}{dt}=-i[H,\rho]+\frac{\kappa}{2}(2a\rho
a^\dag-a^\dag a \rho-\rho a^\dag a)+\frac{\gamma}{2}(2\sigma\rho
\sigma^\dag-\sigma^\dag \sigma \rho-\rho \sigma^\dag \sigma)
\end{equation}

\noindent where $H$ is the system Hamiltonian without any loss as given in Eqn. \ref{eqn:H2}. For the Monte Carlo simulation using Quantum Trajectory method \cite{CarmichaelOpenSystems} the Schr$\ddot{o}$dinger equation is:

\begin{equation}
i\frac{d\psi}{dt}=H_{eff}(t)\psi
\end{equation}

\noindent where $H_{eff}$ is given by:

\begin{equation}
H_{eff}(t)=H(t)-\frac{i}{2}\sum_k D_k^\dag  D_k
\end{equation}

\noindent where, $H(t)$ is the system Hamiltonian without loss (Eqn. \ref{eqn:H2}) and $D_k$ is the collapse operator corresponding to the $k^{th}$ dissipation channel. In the experiment considered here, there are mainly two decay channels: the spontaneous emission of the QD $D_1=\sqrt{\gamma}|g\rangle\langle e|$ and the cavity decay $D_2=\sqrt{\kappa}a$. The dephasing rate of the QD and any non-radiative decay is neglected. As the non-classical state is collected from one of the output modes of the cavity ($b_{out}$), only the collapse operator corresponding to the cavity decay is monitored.

The driving term $\Omega(t)$ in the Hamiltonian described in Eqn. \ref{eqn:H2} is assumed to be of the form $\Omega(t)=\Omega_o p(t)$, where $\Omega_o$ is the amplitude and the time dependence is described by $p(t)$. For the simulation we assume a Gaussian shape for $p(t)$:
 
 \begin{equation}
p(t)=exp\left(-\left(\frac{t-t_{0}}{\tau}\right)^2\right),
\end{equation}

\noindent where $t_{0}$ is the time when the pulse reaches its maximum value and $\tau$ is the pulse-width. 

The non-classical state of light emitted at the output is analyzed using an ideal single photon detector. For each quantum trajectory, a laser pulse is coupled to the cavity and the number of clicks detected at the output is monitored. Ideally, for a deterministic single photon source, a single click should be registered by the detector every time the device is operated. However, the output field is not in a pure single photon state, and in a Fock state basis, it can be expressed as:

\begin{equation}
\ket{b_{out}}=\sum^{\infty}_{n=0} \varphi_{n}\ket{n}
\end{equation}

\noindent where the $\varphi_{n}$ is the coefficient of the Fock state $\ket{n}$. Here we write the output state as a pure state considering that the dephasing rate of the system is negligible. The normalized value ($|c_{n}|^{2}$) of the coefficients $|\varphi_{n}|^{2}$ can be estimated from the number of detected photons at the output when running a large number of trajectories. For example, $|c_{n}|^{2}=\frac{|\varphi_{n}|^{2}}{\sum_{i} |\varphi_{i}|^{2}}$ is well estimated by the relative number of trajectories for which $n$ counts were detected at the output. If the desired output state should be as close as possible to a single photon state, then the simulation parameters should be optimized such that $|c_{1}|^{2}$ is maximized.

The experimental configuration considered here is as shown in Figure \ref{Fig_shaping}(a). The cavity has two mirrors, with decay rates $\kappa_1$ and $\kappa_2$ such that $\kappa_1<<\kappa_2$. Effectively, the total decay rate of the cavity is $\kappa \approx \kappa_1$. The driving laser is incident on the mirror with higher reflectivity and the output field is mainly collected from the lossier mirror. This configuration allows for efficient collection of the non-classical field at the cavity output.

To illustrate the behavior of the system operating in photon blockade under pulsed driving, we first analyze a system with parameters $\kappa/2\pi=1GHz$, $\gamma/2\pi=0.1GHz$, $g/2\pi=40GHz$. The value $\kappa/2\pi=1GHz$ corresponds to a cavity with a quality factor of $Q=160000$. This is about 5 times larger than the state of the art values of $Q$ observed in GaAs cavities with coupled InAs quantum dots operating around 930nm, but still withing the theoretical limit for this material \cite{2009.Arakawa.SCtoLasing}. Regarding the coupling rate $g$, the typical values measured so far are around $g/2\pi=25GHz$\cite{2009.Englund.CoherentExcitation}. However, with further improvements in the material system and the fabrication techniques it is expected that higher values for $Q$ and $g$, as considered here, will be achievable. For this simulation, the cavity and the quantum dot are assumed to be on resonance ($\omega_{c}=\omega_{a}$). The center frequency of the driving field is set on resonance with the transition to the first order manifold ($\omega_{c}+g$) and the pulse width is set to $\tau = 0.45/\kappa$. Figure \ref{Fig_shaping}(b) shows the zero-photon, single photon as well as multi-photon population ($|c_{multi}|^{2}=\sum^{\infty}_{n=2}|c_{n}|^{2}$) as a function of the amplitude of a driving pulse. The values for the coefficients $|c_{n}|^{2}$ are inferred from the number of detection events detected over multiple (3000) quantum trajectories.

As shown in Fig.\ref{Fig_shaping}(b), the probability of obtaining a single photon at the output has a strong oscillatory dependence on the intensity on the incoming pulse. When operating in blockade regime, the successful blocking of the second photon depends on how well the first photon is coupled to the QD-cavity system. One could consider the ground state and the state $\ket{1,-}$ to constitute an effective two-level system. Under pulsed excitation, the population in $\ket{1,-}$ Rabi oscillates with the pulse area(Fig. \ref{Fig_Photon_Blockade}(b)). For optimum operation, one would choose the pulse area such that the system transitions completely from $\ket{0}$ to $\ket{1,-}$ so one photon is coupled into the system and then is released at the output port. At the operation point with maximum single photon state probability, the output field has $\sim 83\%$ single photon, $\sim 1\%$ vacuum state and $\sim 15\%$ multi-photon state character (i.e. $|c_{0}|^{2}=0.01$, $|c_{1}|^{2}=0.83$, $|c_{multi}|^{2}=\sum^{\infty}_{n=2}|c_{n}|^{2}=0.16$).

The oscillations of the single photon population in the output field can also be observed when the maximum field intensity is kept constant and the pulse length is changed. This is shown in Fig.\ref{Fig_shaping}(c), where the pulse intensity was chosen as the intensity that gave the maximum single photon output in Fig.\ref{Fig_shaping}(c) ($\Omega_{0}=10GHz$).

A similar but less prominent effect can be observed in a system with $\kappa/2\pi=5GHz$, $g/2\pi=30GHz$, parameters that are very close to those already achieved experimentally \cite{2009.Englund.CoherentExcitation,2009.Arakawa.SCtoLasing}. For this set of parameters the maximum achievable single photon probability is $\sim 45\%$, with $\sim 30\%$ vacuum state and $\sim 25\%$ multi photon state, as shown in Fig.\ref{Fig_shaping}(d).

One of the main advantages of generating non-classical states of light using photon blockade instead of above band excitation and spectral filtering, is the high degree of indistinguishability of the output field. The two main effects that affect the single photon indistiguishability is the quantum dot jitter and dephasing\cite{Santori02}. While jitter is completely avoided using this method, the output field still suffers from the dephasing of the quantum dot while the optical pulse passes through the cavity. To maximize the indistinguishability, it is thus desirable to operate with shorter laser pulses.

\section{Non-classical state generation via fast control of dipole frequency}
\label{sec:stark_shift_operation}

\begin{figure*}
\centering
\includegraphics[width=5in]{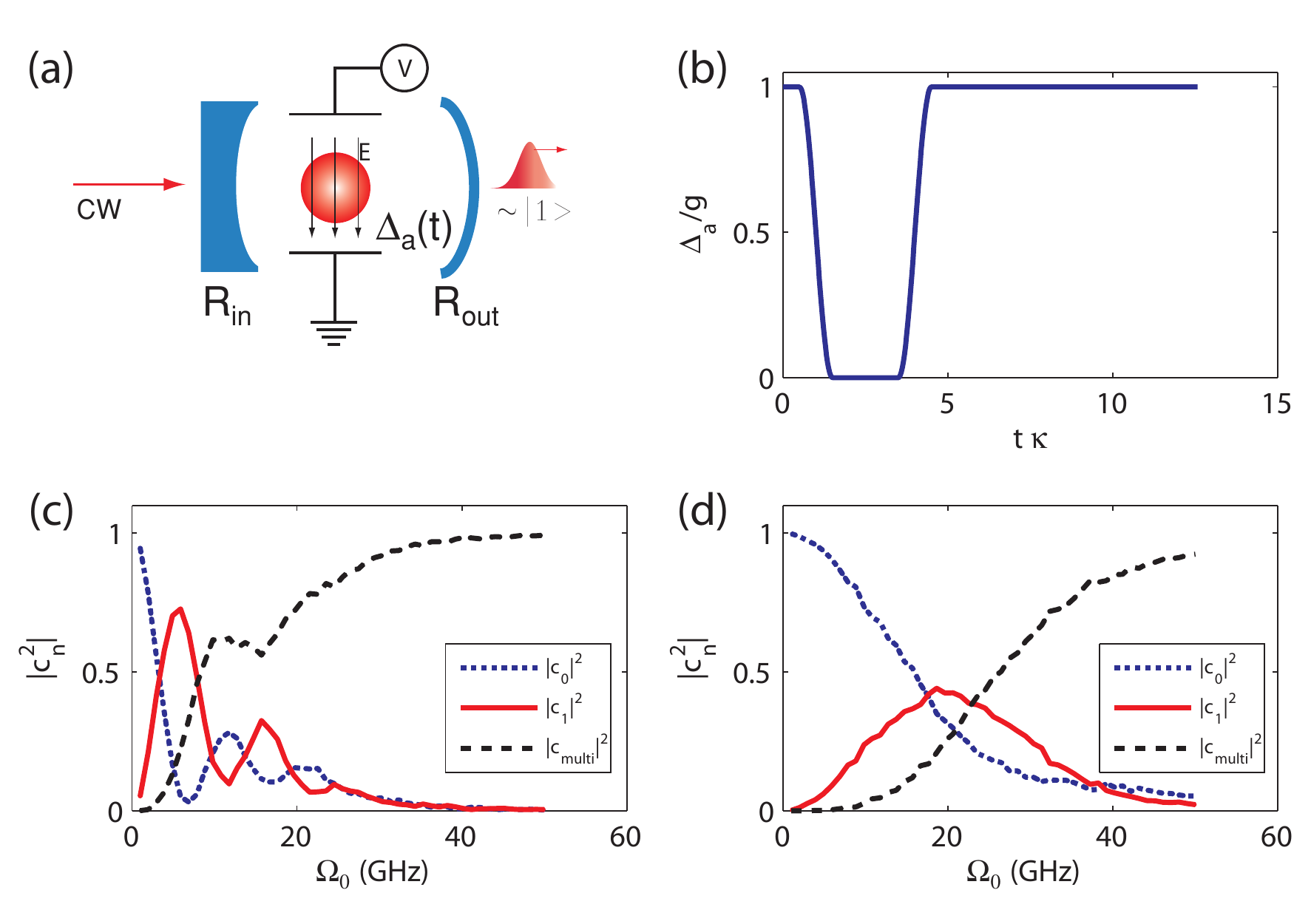}
\caption{(a) Operation principle for non-classical state generation using a continuous wave input and fast control of the dipole frequency using an electric field. (b) Detuning of the dipole frequency with time. The cavity resonance is constant and zero detuning means dipole resonant with the cavity. (c) Normalized Fock-state coefficients ($|c_{0}|^{2}$ for vacuum state, $|c_{1}|^{2}$ for single photon state, and $|c_{multi}|^{2}=\sum^{\infty}_{n=2}|c_{n}|^{2}$ for multi photon probability) for the output field with varying intensity of the continuous wave input field ($\Omega_0$). The system has parameters $\kappa/2\pi=1GHz$, $\gamma/2\pi=0.1GHz$, $g/2\pi=40GHz$. (d) Normalized Fock-state coefficients for a system with $\kappa/2\pi=5GHz$, $\gamma/2\pi=0.1GHz$, $g/2\pi=30GHz$}
\label{electrical_control}
\end{figure*}

In this section we present an alternative method for on demand generation of non-classical light that is based on ultra-fast control of the optical dipole. The principle is depicted in Fig.\ref{electrical_control}(a), where a continuous wave field is incident on the input port and the frequency of the dipole ($\Delta _{a}$) is controlled at time-scales comparable to the cavity decay rate. One method to achieve this kind of ultra-fast control in solid-state systems is by using an electric-field to shift the frequency of a quantum dot via quantum confined Stark effect\cite{2009.Faraon.ElectroOptoSwitch,2009.APL.Petroff.GHzQDControl}. 

In this configuration, the cavity frequency is kept constant and the electric field controlls the cavity - quantum dot detuning. With the quantum dot and the cavity on resonance, the input laser is tuned to the frequency of one of the first order eigenstates (say $\ket{1,-}$). Shifting the resonance of the quantum dot causes an energy shift of both first order eigenstates. To generate single photon states, one should start with the quantum dot detuned from the cavity such that no light is transmitted to the output. Then the quantum dot is brought into resonance with the cavity such that $\ket{1,-}$ becomes resonant with the input laser beam. The QD is kept resonant with the cavity so only one photon is coupled and transmitted through the system, and then it is detuned back as shown in Fig.\ref{electrical_control}(b).

The underlying Hamiltonian for this system is given in RWA by Eq.\ref{eq:hamiltRWA}:

\begin{equation}
\label{eqn:H3} H=\hbar\left[\Delta_{a}(t) \sigma_+ \sigma_-+\Delta_c
a^\dag a+ig(a^\dag\sigma_--a\sigma_+)+\Omega_0(a+a^\dag )\right]
\label{eq:hamiltRWA}
\end{equation}

In this case, the coherent driving term $\Omega_0$ is kept constant, $\Delta_{a}(t)$ is the time dependent detuning of the dipole with the cavity, and $\Delta_{c}$ is the fixed detuning between the cavity and the driving laser.

The quantum statistics of the output field as a function of $\Omega_0$ is shown in Fig. \ref{electrical_control}(c-d) for different parameters of the strongly coupled system. The quantum dot frequency is detuned by up to $g$, as shown in in Fig. \ref{electrical_control}(b). Similar to the optical time domain pulse shaping, oscillations are observed in the magnitude of the single photon state for $\kappa/2\pi=1GHz$, $g/2\pi=40GHz$ (Fig.\ref{electrical_control}(c)). For optimal operation parameters, $|c_{0}|^{2}=0.05$, $|c_{1}|^{2}=0.73$, $\sum^{\infty}_{n=2}|c_{n}|^{2}=0.22$. The results for  $\kappa/2\pi=5GHz$, $g/2\pi=30GHz$ are shown in Fig. \ref{electrical_control}(d), and indicate an $44 \%$ single photon, $35 \%$ vacuum state, and $21 \%$ multi photon Fock states at the point of maxium single photon state probability.

\section{Conclusion}

In conclusion, we have shown that the non-classical states of light that can be generated using the strong optical nonlinearity of a solid-state cavity QED system (photon blockade regime) have a strong dependence on the properties of the input driving field. The parameters of the driving field were optimized such that the main component of the output field is the single photon state. We demonstrate that non-classical states that are $83 \%$ single photon can be obtained using solid state systems with quality factor of $Q=160000$ and coupling rate $g/2\pi = 40 GHz$ which are realistically achievable. These types of non-classical states represent a new tool in the developing toolbox of quantum techonlogies, that could be used effectively to improve the performance of quantum cryptography systems and linear optics quantum computation.

\acknowledgements
This work has been supported by DARPA and NSF. A.M. was also supported by the Stanford Graduate Fellowship (Texas Instruments fellowship). The authors thank Prof. Hideo Mabuchi for useful discussion.

\end{document}